\documentclass[preprint,pteplogo]{ptephy}

\begin{document}

\title{Can we explain AMS-02 antiproton and positron excesses
simultaneously by nearby supernovae without pulsars nor dark matter?}

\author{\name{\fname{Kazunori} \surname{Kohri}}{1,\ast}, \name{\fname{Kunihito} \surname{Ioka}}{1} \name{\fname{Yutaka} \surname{Fujita}}{2}, and \name{\fname{Ryo} \surname{Yamazaki}}{3,4}
}

\address{\affil{1}{Theory Center, IPNS, KEK,  and Sokendai, 1-1 Oho, Tsukuba 305-0801, Japan}
\affil{2}{Department of Earth and Space Science, Graduate School of Science, Osaka  University, Toyonaka, Osaka 560-0043, Japan}
\affil{3}{Department of Physics and  Mathematics, Aoyama-Gakuin University, Kanagawa 252-5258, Japan }
\affil{4}{Harvard-Smithsonian Center for Astrophysics, MS-51, 60 Garden Street, Cambridge, MA 02138, USA}
\email{kohri@post.kek.jp}}

\begin{abstract}
  We explain the excess of the antiproton fraction recently reported
  by the AMS-02 experiment by considering collisions between
  cosmic-ray protons accelerated by a local supernova remnant (SNR)
  and the surrounding dense cloud. The same ``$pp$ collisions''
  provide the right ratio of daughter particles to fit the observed
  positron excess simultaneously in the natural model parameters.  The
  supernova happened in relatively lower metalicity than the major
  cosmic-ray sources.  The cutoff energy of electrons marks the
  supernova age of $\sim 10^{5}$ years, while the antiproton excess
  may extend to higher energy.  Both antiproton and positron fluxes
  are completely consistent with our predictions in Ref.~[4].
\end{abstract}


\maketitle


\section{Introduction}

Recently the Alpha Magnetic Spectrometer (AMS-02) on the International
Space Station has reported that the antiproton to proton ratio stays
constant from 20 GeV to 450 GeV kinetic energy~\cite{AMS-02DATA}.
This behavior cannot be explained by the secondary antiprotons from
collisions of ordinary cosmic rays with interstellar
medium~\cite[e.g.,][]{Kachelriess:2015wpa}.  This suggests a new
source such as astrophysical accelerators and annihilating or decaying
dark matter, although there are still uncertainties in the background
modeling~\cite{Giesen:2015ufa}.

The excess of antiprotons looks surprisingly similar to what we
predicted~\cite{Fujita:2009wk} when the PAMELA experiment detected the
positron excess~\cite{Adriani:2008zr} and the Fermi, HESS, and
ATIC/PPB-BETS experiments observed the electron
anomaly~\cite{cha08,Abdo:2009zk,Aharonian:2009ah}.  We considered
recent supernova explosions in a dense gas cloud (DC) near the Earth.
The antiprotons and positrons are produced as secondaries by the $pp$
collisions between cosmic-ray protons accelerated by the supernova
remnant (SNR) and target protons in a DC which surrounds the
SNRs~\cite{Fujita:2009wk}.  Since the fundamental process determines
the branching fraction, the positron excess should accompany the
antiproton excess.

There are several variants of such a hadronic model, e.g., the
reacceleration of secondaries by SNR shocks~\cite{Blasi:2009bd} or the
non-standard propagation that increases secondaries from ordinary
cosmic-ray collisions with interstellar
matter~\cite{Blum:2013zsa,Cowsik:2013woa,Guo:2014laa}.  
Since the element ratio is the same as the
ordinary cosmic rays in these models, 
the ratio of secondaries (e.g., Li, Be, B) to
primaries (C, N, O) must also rise with energy beyond $\sim 100$ GeV,
which is not
observed yet~\cite{Mertsch:2009ph,Mertsch:2014poa,Cholis:2013lwa}.  In
contrast to these models, our model can accommodate the observations
as shown below.

Before AMS-02, the antiproton observations were consistent with the
secondary background~\cite{Adriani:2008zq}.  Thus the leading models
for the positron excess are leptonic, such as pulsars and leptophillic
dark
matter~\cite[e.g.,][]{Serpico:2011wg,Fan:2010yq,Ioka:2008cv,Kashiyama:2010ui}.
The pulsars cannot usually explain the antiproton excess.  On the
other hand, the dark matter for the positron excess is now severely
constrained by other messengers such as gamma-rays and cosmic
microwave background~\cite[e.g.,][]{Ackermann:2015zua,Ade:2015xua},
and hence we may need fine tunings in dark matter models to reproduce
both the antiproton and positron excesses~\cite{Giesen:2015ufa,DM2015,Evoli:2015vaa,Kohri:2013sva}.

Following Occam's razor, we reexamine our nearby SNR model 
and simultaneously fit the antiproton fraction, positron fraction, 
and total electron and positron flux in light of new AMS-02 data.
In particular the $pp$ collisions give the correct branching fraction
for the observed positron to antiproton ratio.
Throughout this paper we adopt units of $c = \hbar = k_B = 1$.

\section{Supernova explosions in a Dense Cloud}

Here we consider supernova explosions which occurred around $\sim
10^{5} - 10^{6} $ years ago in a  DC. We assume the
DC is located at around $\sim$~100--200~pc away from the Earth 
like the progenitor DCs that produced the Local Bubble (LB) or
Loop~I. In general a massive star tends to be born in a giant
DC~\cite{lar82} which explodes as a supernova. In this paper, we
assume that the Giant DC is ionized and the temperature is
approximately $\sim 10^4$~K at the time of its
explosion~\cite{whi79}. The shock of an SNR accelerates protons, which
produce copious energetic mesons (pions and kaons, etc.) and baryons
(antiproton, proton, antineutron, neutron, etc.)  through the $pp$
collisions in the surrounding DC. The mesons further decay into
energetic positrons, electrons, gamma-rays, and neutrinos.  In total
those local secondary particles can be observed at the Earth as
cosmic-rays in addition to the standard background components.

The energy spectrum of the accelerated protons is parametrized by
\begin{equation}
\label{eq:NE}
 \frac{dn_p}{dE_p} \propto E_p^{-s}  e^{- \frac{E_p}{E_{\rm max, p}}},
\end{equation}
where $s$ is the spectral index. 
The age of SNR, $t_{\rm age}$, approximately determines the maximum
energy~\cite{yam06},
\begin{equation}
 E_{\rm max,p} \sim 2 \times10^2  v_{s,8}^2
\left(\frac{B_{\rm d}}{10~\rm\mu G}\right)
\left(\frac{t_{\rm age}}{10^5{\rm yr}}\right)~{\rm TeV}~,
\label{eq:Emax_p}
\end{equation}
where the shock velocity, $v_s$, is $v_{s,8}=v_s/10^8$cm~s$^{-1} \sim
O(1)$. $B_{\rm d}$ is the downstream magnetic field. We take
the minimum energy of the protons to be its rest mass.  We assume that
the supernova explodes at the center of a DC for simplicity.  In
addition, we also assume that the acceleration stops when the Mach
number of the shock decreases to 7~\cite{Fujita:2009wk}. We define
this time as the acceleration time
$t_{\rm acc}=t_{\rm age}$, 
and the energy spectrum at
this time is given by $s \sim 2$ and $E_{\rm max,p} \sim 120$
TeV~\cite{Fujita:2009wk}. The SNR continues to expand even at $t_{\rm
  age} > t_{\rm acc}$.

The radius is $50$~pc at
$t_{\rm age}=5\times 10^5 $~yr. Since it is comparable to the size of
a giant DC, $R_{\rm DC}$,~\cite{mck07} and the initial energy of the
ejecta from the supernovae is larger than the binding energy of a DC,
the cloud would be destroyed around this time. Until it is destroyed,
the DC is illuminated by the accelerated protons from the inside with
the spectrum of Eq. (\ref{eq:NE}) given at $t_{\rm age}\sim t_{\rm
  acc}$. The duration of the exposure, $t_{pp}$, could be approximated
by the time elapsing from the explosion of the supernovae to the
destruction of the DC  because the timescale $t_{\rm acc}$ is
shorter than $5\times 10^5 $~yr.

After the destruction of the DC, the produced charged particles such
as $\bar{p}$, $p$, $e^{+}$, or $e^{-}$ propagate through diffusion
processes and reach to the Earth. Since we assume that the DC has
already been destroyed well before the present epoch, there are some
differences in arrival times between those charged particles and
massless neutral particles such as photons. It should be a reasonable
assumption that we would not detect any photon and neutrino signals
from the DC $\sim 10^{5-6}$ years after the destruction of the DC.

We have calculated spectra of those daughter particles through the
$pp$ collisions by performing the PYTHIA Monte-Carlo event
generator~\cite{Sjostrand:2006za} (See~\cite{yam06} for the
details). Then we solve the diffusion equation of the charged particle
``$i$'' ($i$ runs $\bar{p}$, $p$, $e^{+}$, and $e^{-}$, ),
\begin{equation}
    \label{eq:diff_eq}
    \frac{\partial f_{i}}{\partial t} 
= K(\varepsilon_{i}) \Delta f_{i} +
    \frac{\partial}{ \partial \varepsilon_{i}} \left[
    B(\varepsilon_{i}) 
f_{i}\right] + Q(\varepsilon_{i})
\end{equation}
where $f_{i}(t,{\boldmath{x},\varepsilon_{i}})$ is the distribution
function of an $i$ particle, and $\varepsilon_{i} = E_{i}/{\rm GeV}$
with $E_{i}$ being the energy of the $i$ particle.  The flux is given
by 
\begin{eqnarray}
  \label{eq:totalflux0}
  \Phi_{i} (t,{\boldmath{x},\varepsilon_{i}}) = \frac{1}{4\pi} f_{i}.
\end{eqnarray}
We adopt a diffusion model 08-005 given in~\cite{Moskalenko:1997gh}
with the diffusion coefficient,
\begin{equation}
 K(\varepsilon_{e}) = K_{0}
 \left(1 +
\frac{\varepsilon_{e}}{3{\rm GeV} }
\right)^{\delta},
\end{equation}
with $K_{0} = 2 \times 10^{28} {\rm cm}^{2}{\rm s}^{-1}$ and $\delta =
0.42$~\cite{AMS-02:2013conf,Genolini:2015cta,Evoli:2015vaa}.  The cooling rate through the synchrotron emission and the
inverse Compton scattering is collectively parametrized to be~\cite{Baltz:1998xv}
\begin{equation}
 B(\varepsilon_{e}) \sim  10^{-16} {\rm s}^{-1} \varepsilon_{e}^{2 }
\left[ 0.2
\left(
\frac{B_{\rm diff} }  {3 \mu {\rm G}}
\right)^{2} + 0.9 
\right],
\end{equation}
where $B_{\rm diff}$ is the magnetic field outside the DC. This set of
the parameters approximately corresponds to the MED model of the
cosmic-ray propagation~\cite{Bottino:2005xy}.

If we assume that the timescale of the production is shorter than that
of the diffusion, $\sim d^{2}/(Kc)$ with $d$ being the distance to the
source, and the source of the daughter particles is spatially
localized sufficiently, we can use the known analytical solution
in~\cite{Atoian:1995ux}.  When the shape of the source spectrum is a
power-law with an index $\alpha$ to be
\begin{eqnarray}
  \label{eq:Qdetail}
  Q =Q_{0}\varepsilon^{-\alpha}\delta(\boldmath{x})\delta(t),
\end{eqnarray}
then the solution is given by
\begin{equation}
    \label{eq:f_diff}
f_{e}=\frac{Q_0 }{\pi^{3/2} d_{\rm diff}^3}
\varepsilon_e^{-\alpha}
\left(1-\frac{\varepsilon_e}{\varepsilon_{\rm cut}}\right)^{\alpha-2}
e^{-(\frac{\bar{d}} {d_{\rm diff}})^2},
\end{equation}
where $\varepsilon_{\rm cut} = \varepsilon_e^2 / B t_{\rm diff}$, and the
diffusion length is represented by
\begin{equation}
 d_{\rm diff} = 2 \sqrt{K t_{\rm diff} \frac{1 - (1 -
\frac{\varepsilon_e} {\varepsilon_{\rm cut}})^{1 - \delta}}{
(1-\delta) 
\frac {\varepsilon_e } {\varepsilon_{\rm cut}} }}\:.
\end{equation}
$\bar{d}$ means the effective distance to the source by spatially
averaging the distance to the volume element of the source, and we
assume $\alpha \simeq s$. We approximately have
\begin{eqnarray}
  \label{eq:Q0emalpha}
  Q_0\varepsilon_{i}^{-\alpha} \sim  V_{s}t_{pp} \frac{d^{2}n_{i}}{dtdE_{i}} 
\end{eqnarray}
with $V_{s}$ the  source volume  where
%
\begin{equation}
 \frac{d^{2}n_{i}}{dtdE_{i}} = \int d E_{p} n_{0} \frac{dn_p}{dE_p}  \sum_{j}
g_{j}  \frac{v_pd\sigma_{j}}{dE_{i}} \:.
\end{equation}
The differential cross section of the ``$j$''-mode for the production
of the $i$ particle is represented to be
$d\sigma_{j}(E_{p},E_{i})/dE_{i}$ with the multiplicity into the
$j$-mode, $g_{j} = g_{j}(E_{p},E_{i})$. $v_p=v_p(E_p)$ is the velocity of the
primary proton.  We also consider the free neutron (antineutron)  decay
for the electron (positron) production process, which is not included
in the original version of PYTHIA. The initial proton spectrum
$\frac{dn_p}{dE_p}$ can be obtained by a normalization to satisfy
\begin{equation}
 V_{s}\int dE_{p} \frac{dn_p}{dE_p}= E_{\rm tot, p} .
\end{equation}
For the local propagation of protons and antiprotons, their cooling is
negligible unlike electrons and positrons.  Additionally we can omit
annihilations of antiprotons through scattering off the background
protons because the scattering rate is small. We can also omit
convection by interstellar turbulence
within the galaxy. An analytical
solution for the proton and the antiproton is also given by the same
equation as Eq.~(\ref{eq:f_diff}) with a limit of
$\varepsilon_p/\varepsilon_{\rm cut} = 0$.

\section{Antiproton and positron fittings}

\begin{figure}[t]
  \begin{center}
    \vspace{-0. cm}
    \includegraphics[width=100mm]{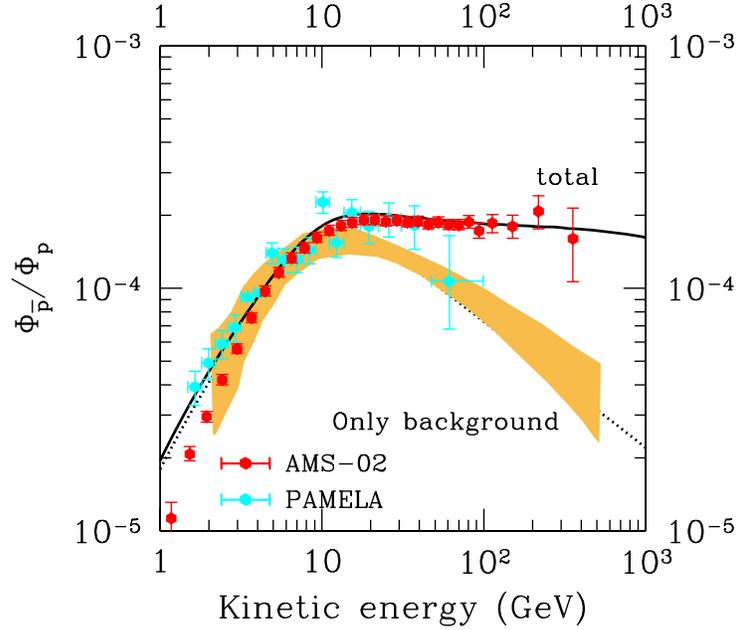}
    \vspace{-1. cm}
    \caption{Antiproton fraction fitted to the data. The data points
      are taken from~\cite{AMS-02DATA} for AMS-02, and
      from~\cite{Adriani:2008zq} for PAMELA. The dotted line is plotted
      only by using the background flux~\cite{Nezri:2009jd}. The
      shadow region represents the uncertainties of the background
      flux among the propagation models shown
      in~\cite{AMS-02DATA}. Cosmic rays below an energy $\lesssim$
      10GeV are affected by the solar modulation. 
        We choose the background line and its uncertainty band only
        for a demonstration purpose. This choice is not
        essential for our conclusion (See the text 
        about Fig.~\ref{fig:posipbar}).} 
  \end{center}
\label{fig:anti_p}
\end{figure}

In Fig.~1, we plot the antiproton fraction at the Earth
in our model (See also a similar model named ``model B'' given in
Ref.~\cite{Fujita:2009wk}). For the background flux, we adopted the
20$\%$ smaller value of the mean value shown
in~\cite{Nezri:2009jd}.  Here, the radius of a spherical DC,
$R_{\rm DC}=40$~pc is adopted. The target proton density is set to be
$n_0 = 50\rm\: cm^{-3}$. The spectral index $s=2.15$ 
and the maximum
energy, $E_{\rm max}=100$~TeV, are assumed. We take the duration of the
$pp$ collision to be $t_{pp}=2\times 10^5$~yr.  The total energy of the
accelerated protons is assumed to be
$E_{\rm tot,p}=2.6 \times 10^{50}$~erg.  The distance to the front of
the DC is set to be $d=200$~pc. About the diffusion time of $e^-$ and
$e^+$, $t_{\rm diff}=2\times 10^5$~yr is adopted. We take the magnetic
field outside the DC to be $B_{\rm diff}=3\rm\: \mu G$
(See~\cite{Fujita:2009wk} for the further details).

\begin{figure}[t]
  \begin{center}
    \includegraphics[width=100mm]{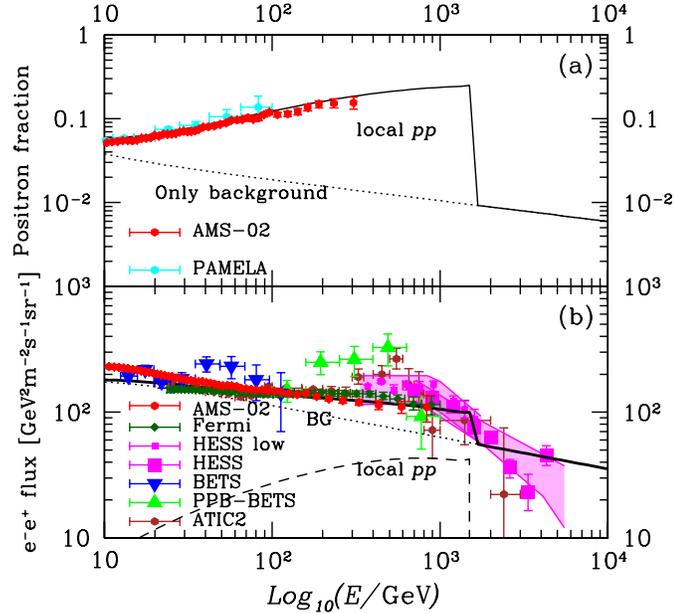} 
    \vspace{-1.3 cm}
    \caption{(a) Positron fraction (solid line), which includes the
      electrons and positrons coming from the DC and background
      electrons (dotted line, for example see
      Refs.~\cite{Moskalenko:1997gh,Baltz:1998xv}). Filled circles
      correspond to the AMS-02 data~\cite{AMS-02DATA,AMSe+e-14,AMSv1}
      and PAMELA data~\cite{Adriani:2008zr} (b) Total electron and
      positron flux (solid line). The flux of the electrons and
      positrons created only in the DC (background) is plotted by the
      dashed (dotted) line. Observational data by AMS-02, Fermi, HESS,
      BETS, PPB-BETS, and
      ATIC2~\cite{cha08,Abdo:2009zk,Aharonian:2009ah,AMSe++e-14} are
      also plotted. The shadow region represents the uncertainty of
      the HESS data.}
  \end{center}
 \label{fig:posifra}
\end{figure}

In Fig.~2, we also plot the positron fraction and the total
$e^-$+$e^+$ flux.  It is remarkable that we can automatically fit the
observational data of both the positron fraction and the total $e^-$ +
$e^+$ flux by using the same set of the
parameters~\cite{Fujita:2009wk}. Here the cooling cutoff energy is
approximately given by $\varepsilon_{\rm cut} = \epsilon_e^2  / B t_{\rm diff}
\sim$~1~TeV ($t_{\rm diff} / 2 \times 10^5$ yrs)$^{-1}$.

The positron fraction rises at higher energies than that of the
antiproton fraction (Fig.~\ref{fig:posifra}), because the spectral
index of the background antiproton is harder than that of the
background positron.  This comes from a difference between their
cooling processes.  Only for background positrons and electrons
the cooling is effective in the current situation.

\begin{figure}[htbp]
  \begin{center}
    \includegraphics[width=100mm]{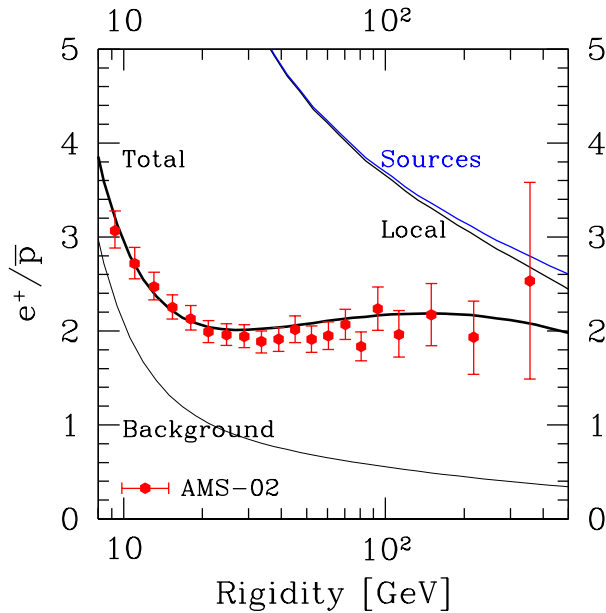} 
    \vspace{-1.1 cm}
    \caption{Positron to antiproton ratio as a function of the
      rigidity with adding the local components produced by the $pp$
      collisions occurred at SNRs near the Earth. The thick solid line
      represents the case of the total flux. From the upper right to
      the lower left, we plot the flux ratios of 1) the one at the
      source (without cooling), 2) only the local components, 3) the
      total of the local and the background components, and 4) only
      the background components. The observational data reported by
      AMS-02 are also plotted. }
  \end{center}
 \label{fig:posipbar}
\end{figure}

In Fig.~\ref{fig:posipbar}, we plot the positron to antiproton ratio
as a function of the rigidity. Here the local components represent the
contribution of the nearby SNRs produced only by the $pp$ collisions.
From this figure, we find that both of the positron and the
antiproton can be consistently fitted only by adding astrophysical
local contributions produced from the same $pp$ collision sources.

\section{Conclusion}
We have discussed the anomaly of the antiproton fraction
recently-reported by the AMS-02 experiment. By considering the same
origin of the $pp$ collisions between cosmic-ray protons accelerated
by SNRs and a dense cloud which surrounds the SNRs, we can fit the
data of the observed antiproton and positron simultaneously in the
natural model parameters. The observed fluxes of both antiprotons and
positrons are consistent with our predictions shown in
Ref.~\cite{Fujita:2009wk}.

Regardless of the model details, the ratio of antiproton to positron
is essentially determined by the fundamental branching fraction into
each mode of the $pp$ collisions.  Thus the observed antiproton excess
should entail the positron excess, and vice versa.  This does not
depend on the propagation model since both antiparticles propagate in
a similar way below the cooling cutoff energy $\sim$ TeV.

The cutoff energy of $e^-$ cooling marks the supernova age of
$\sim 10^{5}$ years~\cite{Ioka:2008cv,Kawanaka:2009dk}, while we also
expect a $e^{+}$ cutoff.  The trans-TeV energy will be probed by the
future CALET, DAMPE and CTA experiments
\cite{Kobayashi:2003kp,Kawanaka:2010uj}.  An anisotropy of the arrival
direction is also a unique signature, e.g., \cite{Linden:2013mqa}. We
may estimate the amplitude of anisotropy as
$\delta_e \sim 3d/2ct_{\rm diff} \sim 0.5\%$, which is below the upper
limits by Fermi observations~\cite{Ackermann:2010}.

The boron to carbon ratio as well as the Li to carbon ratio have no
clear excesses~\cite{AMS-02DATA}.  This suggests that the carbon
fraction of the excess-making cosmic rays is smaller than that of the
ordinary cosmic rays.  In general the supernovae in the DC would not
be the main channel of cosmic-ray production.  Most of cosmic rays 
above $\sim 30$ GeV may
be produced in chemically enriched regions, such as superbubbles, as
implied by the hard spectrum of cosmic-ray
helium~\cite{Ohira:2010eq}. Or the carbon abundance of the destroyed
DC might happen to be lower than the Galactic
average~\cite{Fujita:2009wk}.

We should be careful about the background systematics.  In particular
the propagation uncertainties yield the largest
errors~\cite{Yuan:2014pka,Giesen:2015ufa}.  However, in the energy
region above $\sim 100$ GeV where the background contribution is
small, the observed positron to antiproton ratio is very close to the
branching fraction of the $pp$ collisions (source components in
Fig.~\ref{fig:posipbar}).
 This fact is free from the background choice and partially supports our model.

\section*{ACKNOWLEDGMENTS}
This work was supported in part by Grant-in-Aid for Scientific
research from the Ministry of Education, Science, Sports, and Culture
(MEXT), Japan, Nos. 26105520, 15H05889 (K.K.), 26247042 (K.K. and K.I.),
26287051, 24103006, 24000004 (K.I.), 15K05080 (Y.F.), and 15K05088
(R.Y.).  The work of K.K. and K.I. was also supported by the Center
for the Promotion of Integrated Science (CPIS) of Sokendai
(1HB5804100).

\section*{Note added}
While finalizing this manuscript, Ref.~\cite{Kachelriess:2015oua}
appeared which has some overlaps with this work.


\end{document}